\documentclass[aps,amsmath,article,11pt]{revtex4-2}

\usepackage{amssymb}%

\usepackage{longtable}% Long tables

\usepackage{graphicx}% Include figure files
\usepackage{epstopdf}
\usepackage{dcolumn}% Align table columns on decimal point
\usepackage[percent]{overpic}  % insert LaTex text over figures
\usepackage{bm}
\usepackage{xcolor,xspace}
\usepackage{braket}
\usepackage{wasysym}  % for the use of \CIRCLE
\usepackage[ulem=normalem]{changes}
\usepackage{hyperref}
\hypersetup{colorlinks=True,urlcolor=blue,linkcolor=blue,citecolor=blue,filecolor=black}

\colorlet{Changes@Color}{red}  % changes in red color

\begin{document}

\title{In Memoriam: Igal Talmi (1925-2026)}
\maketitle

\vspace{-12pt}
\begin{figure}[h]
\includegraphics[width=0.45\linewidth]{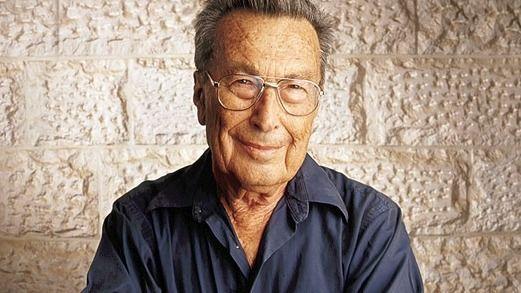}
\end{figure}

\vspace{12pt}
Igal Talmi, a recognized leader in nuclear structure
physics, passed away on 4 February 2026, at the age of
101. Talmi completed his M.Sc. studies in physics
at the Hebrew University of Jerusalem in 1947 with
Giulio Racah. He received the degree of Dr. Sc. Nat.
at the ETH Zurich in 1952 with Wolfgang Pauli. From
1952 to 1954, he was a research fellow at Princeton
University with Eugene Wigner. He joined the Weizmann
Institute of Science in 1954, becoming a professor of
physics in 1958, head of the Department of Nuclear
Physics (1967-1976), and dean of the Faculty of Physics
(1970-1984). Talmi was a member of the Israel Academy
of Sciences and Humanities since 1963. 

Talmi started his scientific career in the early years
of the nuclear shell model. He developed this model and 
brought it to the stage in which the structure of light
nuclei could be understood. In later years, his ideas
provided insights relevant to the microscopic foundations
of the interacting boson model (IBM) of medium-heavy
nuclei. A sample of seminal and lasting contributions
is outlined below.

In his 1952 paper~\cite{talmi52} Talmi introduced
a transformation between products of harmonic oscillator
wave functions, which allows the explicit evaluation of
matrix elements for a given form of two-body interaction
in terms of a finite set of ``Talmi integrals.'' An early
application of this result~\cite{talmi54} revealed the
role of the tensor force in explaining the very long
lifetime of $^{14}$C. Talmi proposed to extract the matrix
elements of the effective interaction in nuclei from
measured energies. Restriction to two-body interactions
reduces the number of matrix elements and implies
relations between different spectra. A first example of
this approach was presented in a 1956 paper [3], which
related the spectra of $^{40}$K and $^{38}$Cl in good
agreement with experiment. Racah viewed this work as
``the beginning of nuclear spectroscopy.''
A 1960 paper~\cite{talmi60} predicted a ground state spin
of $1/2^+$ in $^{11}$Be instead of the expected $1/2^-$,
reflecting a change in the order of neutron $2s_{1/2}$ and
$1p_{1/2}$ levels due to the proton-neutron interaction.
In today's terminology, this experimentally confirmed
scenario exemplifies the disappearance of the N=8 magic
number in $^{11}$Be.

A central theme in Talmi's research is the topic of
pairing and seniority. A work started in 1952 with
Racah~\cite{racah52} culminated in the empirically tested
``Talmi mass formulas'' for $j^n$
configurations~\cite{talmi62}. In 1971, Talmi introduced
``generalized seniority''~\cite{talmi71} a partially
solvable scheme relevant to semi-magic nuclei, based on
monopole and quadrupole pairs of identical valence
nucleons occupying several non-degenerate $j$-orbits.
These favored pairs were used by Talmi and his
colleagues~\cite{talmi77,talmi78} to provide a shell-model
interpretation for the proton-neutron version of the IBM.

In addition to his influential papers, Talmi authored
two books that inspired generations of nuclear structure
physicists: \emph{Nuclear Shell Theory}~\cite{book1},
coauthored with Amos de-Shalit, served as the ``bible''
for users of the shell model, and
\emph{Simple Models of Complex Nuclei}~\cite{book2},
on the shell model and IBM, emphasizes Talmi's quest
for simple and clear explanations of physical phenomena. 
 
In recognition of his achievements, Talmi was awarded
numerous prizes, including the Israel Prize in 1965
(with Amos de-Shalit) and the American Physical Society
Hans Bethe Prize in 2000, cited for ``pioneering work on
the shell model of the nucleus that laid the foundation
of much of what we know about nuclear structure.''  

Igal enjoyed physics discussions, sharing his immense
knowledge, providing guidance and inspiration to many,
young and old. He had many interests beyond physics,
notably nature and bird watching.
He will be greatly missed by all who had
the privilege to know him.

\vspace{14pt}
\begin{flushright}
  Amiram Leviatan and Avraham Gal\\
  The Hebrew University of Jerusalem, Israel
\end{flushright}

\end{document}